\documentclass[prb,amsmath,amssymb,twocolumn,showpacs,floatfix]{revtex4}
\usepackage{graphicx}
\usepackage[english]{babel}
\usepackage{times}
\usepackage{dcolumn}
\usepackage[usenames,dvipsnames]{color}
\usepackage{units}

\begin{document}

\title{Detection of Majorana edge states in topological
superconductors through the non-Fermi-liquid effects induced in an
interacting quantum dot}

\author{Rok \v{Z}itko}

\affiliation{Jo\v{z}ef Stefan Institute, Jamova 39, SI-1000 Ljubljana,
Slovenia,\\
Faculty  of Mathematics and Physics, University of Ljubljana,
Jadranska 19, SI-1000 Ljubljana, Slovenia}

\date{\today}

\begin{abstract}
It is shown that the presence of the continuum of Majorana fermion
edge states along the perimeter of a chiral topological superconductor
can be probed using an interacting quantum dot coupled to three
terminals: the lead supporting the Majorana edge states and two
spin-polarized (ferromagnetic) measurement leads. The hybridization
with the Majorana states induces a particular type of the Kondo effect
with non-Fermi-liquid properties which can be detected by performing
linear conductance measurements between the source and drain
measurement leads: the temperature and magnetic-field dependence of
the conductance is characteristically different from that in the
conventional Kondo effect.
\end{abstract}


\maketitle

\newcommand{\vc}[1]{{\mathbf{#1}}}
\newcommand{\vck}{\vc{k}}
\newcommand{\braket}[2]{\langle#1|#2\rangle}
\newcommand{\expv}[1]{\langle #1 \rangle}
\newcommand{\ket}[1]{| #1 \rangle}
\newcommand{\Tr}{\mathrm{Tr}}
\newcommand{\kor}[1]{\langle\langle #1 \rangle\rangle}
\newcommand{\degg}{^\circ}
\newcommand{\Imm}{\mathrm{Im}}

\section{Introduction}

Two-dimensional (2D) electron systems with gapped bulk states and
gapless edge states have been intensely studied ever since the
discovery of the quantum Hall effect \cite{klitzing1980} and the
emergence of the theories which brought to light the topologically
non-trivial nature of the quantum Hall state \cite{thouless1982}. In
recent years, this line of research has significantly intensified with
the prediction and the subsequent experimental discovery of the
time-reversal-invariant generalizations of the quantum Hall state
where the role of the external magnetic field is played by the strong
spin-orbit coupling \cite{kane2005f, fu2006f, bernevig2006f,
moore2007f, bernevig2006sc, konig2007d}. These systems, now known as
the ``two-dimensional topological insulators'', are insulating in the
bulk but support helical edge states, i.e., a pair of one-dimensional
propagating modes connected by the time-reversal symmetry (Kramers'
pairs) and propagating in the opposite directions for the opposite
(pseudo)spins \cite{bernevig2006f}. The edge states have dispersion
along the edge, but they are confined along the direction
perpendicular to the edge. These states are robust against
perturbations which preserve the time-reversal (TR) invariance, since
their presence is guaranteed by the non-trivial topological properties
of the bulk states. In addition to 2D topological insulators (TI),
there are also three-dimensional TIs with insulating bulk states and
topologically protected gapless chiral surface states, which have
Dirac spectrum. Such materials are also known as ``strong topological
insulators''.

In a metal with a Dirac spectrum Majorana fermion bound states can be
induced by the s-wave superconductivity through the proximity effect
\cite{bergman2009,nilsson2008}. Majorana fermions can be described as
real fermions ($\eta^\dag=\eta$) and have half the degrees of freedom
as the complex Dirac fermions. In other words, a set of fermionic
creation and annihilation operators can be rewritten using a pair of
Majorana operators as $\psi=(\eta_1+i \eta_2)/\sqrt{2}$ and
$\psi^\dag=(\eta_1-i\eta_2)/\sqrt{2}$. This is more than a simple
change of basis, since Majorana states may be spatially separated.
Especially important are the situations where the Majorana modes have
zero energy, as this implies the degeneracy of the ground state
\cite{ivanov2001} and it allows the system to support excitations with
non-Abelian statistics (i.e., particles which are neither fermions nor
bosons) \cite{stern2010}. Such systems would allow reliable non-local
storage of quantum information \cite{kitaev2001} and they would
provide the building blocks for topological quantum computers
\cite{kitaev2003}.  While the non-Abelian states of matter have not
been observed yet, there is now an intensive search for Majorana
excitations in various condensed-matter systems
\cite{wilczek2009,stern2010}.

As the dispersion of the surface-state electrons on a strong TI forms
a Dirac cone, an interesting state has been predicted to emerge by
bringing in contact a TI with an (s-wave) superconductor
\cite{fu2008majorana}. A linear junctions between a superconductor and
a magnet in contact with a TI may namely form a one-dimensional wire
for Majorana fermions \cite{fu2008majorana,fu2009inter}. Such a
``Majorana quantum wire'' can be described as ``half a regular 1D
Fermi gas'' \cite{fu2008majorana}. A number of related systems may
also support Majorana edge modes: regular semiconductors with
spin-orbit coupling in proximity to a superconductor and a magnetic
insulator \cite{sau2010,sau2010prb}, edge states of 2D TIs
\cite{fu2009josephson}, junctions with ferromagnetic insulators
\cite{tanaka2009nfis}, etc.

Unfortunately, Majorana fermions are by their very nature rather
elusive and it is difficult to assert their existence in a given
system. Majorana fermions in superconductors are electrically neutral
and do not couple to external fields. One approach for their detection
has been, for example, to combine two Majorana fermions into a single
Dirac fermion in order to allow probing with charge transport
\cite{fu2009inter, akhmerov2009inter}.  Various detection schemes have
already been proposed for Majorana modes in topological insulators.
Some of them are only capable of detecting the presence of Majorana
modes (either single localized levels or continua of propagating
modes), while others can actually measure the state of the system
(they are, thus, read-out schemes) and could be used to demonstrate
the non-Abelian statistics associated with the Majorana zero-energy
modes. The detection schemes are based on the detection via the
Josephson current \cite{fu2009josephson}, on interferometry
\cite{fu2009inter,akhmerov2009inter,law2009,sau2010pp},
``teleportation'' (non-local electron transfer process which maintains
the phase coherence) \cite{fu2010tele}, flux qubit interferometry
\cite{hassler2010}, or noise measurements \cite{nilsson2008}.

\begin{figure}[htbp!]
\centering
\includegraphics[clip,width=5cm]{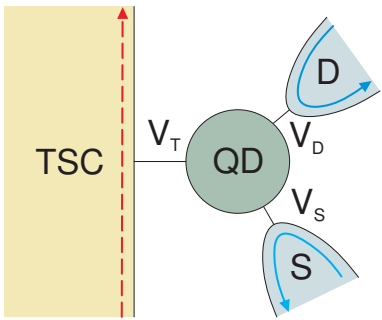}
\caption{(Color online) An interacting quantum dot QD is coupled to a
spin-up Majorana fermion edge channel of a chiral topological
superconductor TSC with hopping $V_T$ and to two ferromagnetic probe
leads (source S and drain D electrodes) which are fully spin polarized
in the spin-down direction with hoppings $V_S$ and $V_D$. Instead of
ferromagnetic probe leads, one may also use two systems in the quantum
anomalous Hall state which have fully spin-polarized edge states. The
hybridization with the Majorana modes localized along the perimeter of
the topological superconductor will induce a non-Fermi-liquid Kondo
effect which can be probed using the ferromagnetic contacts by
measuring the temperature dependence of the (spin-down) linear
conductance through the dot. To tune the system to the
non-Fermi-liquid point, one can change the gate voltage on the dot,
apply an external magnetic field, and change the coupling constants
$V_S$, $V_D$ and $V_T$. Dashed line denotes Majorana electrons, full
line indicates Dirac electrons, while the shades of gray (colors online)
distinguish spin-up and spin-down polarization.}
\label{fig0}
\end{figure}

In this paper a further Majorana mode detection scheme is described. 
It is a simple detection scheme, not a read-out scheme.  It makes use
of the effect of the Majorana modes on the screening of the impurity
spin if an interacting quantum dot is coupled to the Majorana quantum
wire on one side and to two normal (but spin-polarized) measurement
wires on the other side, as shown in Fig.~\ref{fig0}. The idea here is
that Majorana fermions and the non-Fermi-liquid variants of the Kondo
effect (for instance, the two-channel Kondo effect, which is relevant
here) go hand in hand. It will be shown that the coupling of the
quantum dot to an additional Majorana mode will modify the transport
properties of the quantum dot probed by the additional leads. In
particular, it will change the temperature dependence of the linear
conductance. Some aspects of the proposed scheme are related to the
work on quantum dots coupled to the edge states of the $\nu=5/2$
fractional quantum Hall effect (FQHE) \cite{fiete2008, fiete2010,
sevier2011}. The two cases differ in the origin of the degrees of
freedom which are necessary (in addition to the Majorana modes) to
generate the two-channel Kondo effect: in the FQHE, they are the
bosonic edge states [field $\phi$ in Eq.~(1) in
Ref.~\onlinecite{fiete2008}], while here we make use of the
spin-polarized probing leads. The two cases also differ in the
measurement scheme: in the FQHE case, one measures the charge
susceptibility of the dot using capacitively coupled probes, while
here we propose to perform transport experiments.

The description of the generalized Kondo problems with
non-Fermi-liquid fixed points in terms of Majorana modes has been very
fruitful and it allows for a simple interpretation of the finite-size
excitation spectra \cite{maldacena1997,bulla1997sigmatau}. A well
known example is the two-channel Kondo (2CK) effect which has been
intensively discussed theoretically
\cite{nozieres1980,sacramento1991,pang1991,affleck1992,emery1992,sengupta1994,andrei1995,
ye1996,cox1998,zarand2000,zarand2002} and was recently experimentally
realized using semiconductor quantum dots \cite{potok2007}. This type
of the Kondo effect occurs when a single spin-$1/2$ quantum impurity
is equally coupled to two independent screening channels (there must
be no charge transfer between the channels \cite{flnfl3,zarand2006});
this leads to an overscreening effect in which the localized spin
forms a new spin-$1/2$ state by coupling to two neighboring spins from
the leads, this new spin-$1/2$ collective state is then coupled to the
two next-nearest neighbor spins from the leads into another spin-$1/2$
state, and so forth, generating a complex non-local screening cloud
state. The non-Fermi-liquid state associated with the two-channel
Kondo effect can be described using conformal field theories in which
an odd number of Majorana modes have their boundary conditions twisted
due to the presence of the magnetic impurity \cite{maldacena1997}.

Quantum impurities (either in the form of quantum dots or magnetic
impurity atoms) in contact with topological insulators have already
been studied in different contexts. The Kondo effect due to a magnetic
impurity in the helical edge liquid may be affected by the
interactions in the one-dimensional chiral channel \cite{maciejko2009},
although the experiments indicate that the interactions in known
systems appear to be rather weak, with a Luttinger parameter $K\sim1$
\cite{konigphd,maciejko2009}. For a quantum dot coupled to two helical
edge states a variant of the two-channel Kondo effect may occur
\cite{law2010}. It has also been shown that a quantum impurity coupled
to Majorana edge fermions \cite{shindou2010} may be mapped to a
two-level system with Ohmic dissipation. This last problem is somewhat
related to the one discussed in this work, but there is crucial
difference: the model proposed here allows particle exchange with the
Majorana wire, while the model studied in
Ref.~\onlinecite{shindou2010} considers only the exchange coupling
(without discussing its microscopic origin). As commonly observed in
other impurity problems, an exchange-only effective model may behave
rather differently than a model with hopping terms; this appears to be
the case here, too.

We discuss a quantum dot coupled to a Majorana channel on one side and
two spin-polarized leads on the other. When the spin-polarization of
the probe leads is opposite to that of the Majorana channel, the
impurity couples to three Majorana modes, while the fourth mode of the
full Anderson impurity model is absent (or fully decoupled). It has to
be emphasized that the spin-polarized leads are not only ``probe''
leads to measure the transport properties of the system, but they are
crucial for the emergence of the (two-channel) Kondo effect, i.e.,
they participate in the formation of the Kondo state. The full details
of the model considered will be presented in Sec.~\ref{sec2}, where
the numerical techniques will also be briefly described. The results
of numerical calculations will be given in Sec.~\ref{sec3} and we
conclude with a brief discussion of the possible issues in the
experimental realization of the proposed scheme. In an Appendix, we
solve exactly the non-interacting resonant-level model with different
couplings to the Majorana modes of a single conductance channel.

\section{Model and method}
\label{sec2}

For definiteness, we consider the physical realization of a system
supporting a one-dimensional Majorana edge channel as proposed in
Ref.~\onlinecite{qi2010tsc}. The system is a hybrid device made of an
insulator layer in the quantum anomalous Hall (QAH) state and a fully
gapped superconducting layer. Its phase diagram supports a chiral
topological superconductor (TS) phase with an odd number of chiral
Majorana edge modes \cite{qi2010tsc}. The QAH state can be induced by
magnetic doping of topological insulators
\cite{liu2008qah,yu2010qah,qi2010tsc}: as the magnetization increases,
the spin down (for example) edge states penetrate deeper in the bulk
until they disappear by merging with the bulk states, while the spin
up edge states remain bound to the edge. The QAH system thus has
spin-polarized single chiral edge states. When this system then
experiences the superconducting proximity effect, it may be tuned to
become a TS \cite{qi2010tsc}. The single edge mode decomposes into two
chiral Majorana edge modes, one of which penetrates deeper in the bulk
and disappears, while the remaining one persists bound to the edge
\cite{qi2010tsc}. We are thus left with a single spin-polarized (we
choose it as spin-up) chiral Majorana fermion edge state, with the
effective Hamiltonian
\begin{equation}
H_\mathrm{edge} = \sum_{p>0} v p \left( \eta_{-p} \eta_p \right),
\end{equation}
where $v$ is the Fermi velocity, $p$ the momentum, and $\eta_p$ the
Majorana fermion operators which satisfy the canonical Majorana
anticommutation rules $\{ \eta_{p}, \eta_{p'} \}=\delta_{p,p'}$.

The quantum dot is described as a single impurity level $d$:
\begin{equation}
H_\mathrm{dot} = \sum_\sigma \epsilon n_\sigma 
+ U n_\uparrow n_\downarrow 
+ g\mu_B B \frac{1}{2}\left( n_\uparrow - n_\downarrow \right).
\end{equation}
Here $n_\sigma=d^\dag_\sigma d_\sigma$ is the spin-$\sigma$ occupancy
operator, the energy level $\epsilon$ can be controlled by the gate
voltage, $U$ is the on-site charge repulsion, $g$ is the gyromagnetic
ratio, $\mu_B$ the Bohr magneton, and $B$ the external magnetic field.
The dot is coupled to two ferromagnetic leads (assumed to be fully
spin-polarized in the opposite direction compared to the edge states
of the TSC) with parallel alignment of the magnetization in both
leads. The hybridization with these two leads can then be described as
\begin{equation}
\label{h1}
H_1 = \sum_{k,a=\{S,D\}} V_{k} \left( c^\dag_{a,k\downarrow}
d_\downarrow + \text{H.c.} \right),
\end{equation}
where $a$ denotes the lead (source and drain) and $k$ is the momentum,
and $c^\dag_{a,k\sigma}$ is the creation operator for electrons in the
leads. The total hybridization for spin-down electrons can be
characterized by a single quantity $\Gamma_\downarrow = \sum_a \pi
\rho_a |V_{a,k_F}|^2$, where $k_F$ is the Fermi momentum. It should be
noted that the dot couples only with a definite combination of
modes in both leads, thus there is effectively a single channel of
spin-down electrons.

We now consider the coupling of the quantum dot to the edge of the
TSC. The microscopic Hamiltonian in principle takes the form analogous
to Eq.~\eqref{h1}:
\begin{equation}
H_2 = \sum_k V_{T,k} \left( f^\dag_{k\uparrow} d_\uparrow +
\text{H.c.} \right),
\end{equation}
since the electrons which tunnel are true (Dirac) electrons.
Nevertheless, in vicinity of the Fermi level, i.e., inside the gap of
the TSC, the only propagating modes are the Majorana fermions, thus
the operators $f^\dag_{k\uparrow}$ and $f_{k\uparrow}$ are not
independent, but may be expressed in terms of the Majorana operators
$\eta_p$. The impurity level $d$ thus hybridizes only with the
$\eta_p$ Majorana modes which have half the degrees of freedom of the
regular Dirac electrons.

To make the discussion more general, we will nevertheless consider
both Majorana modes which constitute the full complex Dirac electron
(we name them $\eta_1$ and $\eta_2$), so that
\begin{equation}
f^\dag_\uparrow = \left( \eta_1 + i \eta_2 \right)/\sqrt{2},
\quad
f_\uparrow = \left( \eta_1 - i \eta_2 \right)/\sqrt{2},
\end{equation}
but we will allow for different hybridization of $\eta_1$ and
$\eta_2$. We decompose the hopping term as
\begin{equation}
f^\dag_\uparrow d_\uparrow+d^\dag_\uparrow f_\uparrow
= \frac{1}{\sqrt{2}} \left( (\eta_1+i\eta_2) d_\uparrow
+ d^\dag_\uparrow (\eta_1-i\eta_2) \right)
\end{equation}
and introduce separate couplings $t_1$ and $t_2$ for the
two modes:
\begin{equation}
\frac{1}{\sqrt{2}} \left( t_1 (\eta_1 d_\uparrow+d^\dag_\uparrow \eta_1)
+ t_2 (i \eta_2 d_\uparrow - i d^\dag_\uparrow \eta_2) \right).
\end{equation}
The different hybridizations correspond to different spatial
localization of the Majorana modes as the QAH state makes the
transition to the TSC state and one of the two modes penetrates deeper
into the bulk. We then rewrite $\eta_1$ and $\eta_2$ in terms of the
original Dirac operators and find that the coupling Hamiltonian is
proportional to (see also Ref.~\onlinecite{bolech2007})
\begin{equation}
\label{hyb}
V \left(f^\dag_{\uparrow} d_\uparrow + \text{h.c.} \right) +
A \left(f^\dag_{\uparrow} d^\dag_\uparrow + \text{h.c.} \right) 
\end{equation}
where
\begin{equation}
\begin{split}
V = (t_1+t_2)/2,\\
A = (t_1-t_2)/2.
\end{split}
\end{equation}
The limit $t_1=t_2$ ($A=0$) corresponds to the QAH state, while the
limit $t_1 \neq 0$, $t_2=0$ ($V=A$) describes the coupling of the
quantum dot to the edge states of a system in the TSC state. In the
following it will be shown that as $t_2$ is reduced starting from the
initial value of $t_1$, the system makes a transition from the regular
Kondo regime to a non-Fermi-liquid regime with $\ln2/2$ residual
impurity entropy.

The impurity model considered is very closely related to the O(3)
symmetric Anderson model
\cite{coleman1995prl,coleman1995,bulla1997sigmatau,bradley1999} which
has been proposed to study some aspects of the two-channel Kondo model
fixed point. The idea in the cited works is to couple the same spin
impurity degree of freedom to both spin and isospin degrees of freedom
of the same conduction channel, which takes into account the property
of the spin-charge separation in one-dimensional systems. The isospin
degree of freedom (also known as the axial charge or the particle-hole
degree of freedom \cite{jones1988}) for some orbital $d$ is defined by
the operators
\begin{equation}
\begin{split}
I_x &= \frac{1}{2} \left( d^\dag_\uparrow d^\dag_\downarrow + 
d_\downarrow d_\uparrow \right), \\
I_y &= \frac{1}{2} \left( -i d^\dag_\uparrow d^\dag_\downarrow +
i d_\downarrow d_\uparrow \right), \\
I_z &= \frac{1}{2} \left( d^\dag_\uparrow d_\uparrow 
+ d^\dag_\downarrow d_\downarrow - 1 \right),
\end{split}
\end{equation}
which fulfill the SU(2) relations $[I_i,I_j]=i \epsilon^{ijk} I_k$,
just like the spin operators. In other words, a single channel
provides two sets of SU(2) degrees of freedom, associated with charge
and spin, respectively, which become separated on low energy scales.
The spin-isospin Kondo model is then defined as \cite{coleman1995prl}
\begin{equation}
H=H_\mathrm{band} + \left[ J_1 \boldsymbol{\sigma} + J_2
\boldsymbol{\tau} \right] \cdot \vc{S},
\end{equation}
where
\begin{equation}
\begin{split}
\boldsymbol{\sigma} &= \left( \psi^\dag_\uparrow, \psi^\dag_\downarrow
\right) \cdot \left(\frac{1}{2} {\vec{\sigma}} \right)
\cdot \begin{pmatrix} \psi_\uparrow \\ \psi_\downarrow \end{pmatrix},\\
\boldsymbol{\tau} &= \left( \psi^\dag_\uparrow, \psi_\downarrow
\right) \cdot \left(\frac{1}{2} {\vec{\sigma}} \right)
\cdot \begin{pmatrix} \psi_\uparrow \\ \psi^\dag_\downarrow \end{pmatrix}.
\end{split}
\end{equation}
Here $\psi_\sigma^\dag$ is the particle creation operator at the
position of the impurity, $\vec{\sigma}$ is the vector of Pauli
matrices, and $\vc{S}$ is the impurity spin operator. The O(3)
symmetric Anderson model is a variation of the standard symmetric
Anderson model \cite{coleman1995,bulla1997sigmatau}
\begin{equation}
\begin{split}
H &= H_\mathrm{band} + \sum_\sigma V_1 \left( \psi_{\sigma}^\dag d_\sigma
+ \text{H.c.} \right) \\
&+ U \left( n_\uparrow-1/2 \right)
\left( n_\downarrow-1/2 \right),
\end{split}
\end{equation}
with an additional anomalous hybridization term
\begin{equation}
H'=-V_2 \left( d^\dag_\downarrow \psi_{\downarrow}^\dag +
\psi_{\downarrow} d_\downarrow + d^\dag_{\downarrow} \psi_{\downarrow}
+ \psi^\dag_{\downarrow} d_\downarrow \right).
\end{equation}
This model maps to the spin-isospin Kondo model via a Schrieffer-Wolff
transformation \cite{coleman1995,bulla1997sigmatau} with
$J_1=4V_1(V_1-V_2)/U$ and $J_2=4V_1V_2/U$. The parameter $V_2$ in
Refs.~\cite{coleman1995,bulla1997sigmatau} is essentially equivalent
to the parameter $A$ in Eq.~\eqref{hyb}. In particular, the special
point $V=A$ corresponds to the special point $V_2=V_1/2$.

The relation of the 2CK model to the Majorana modes also plays an
important role in the bosonisation and refermionisation approach
by Emery and Kivelson who have shown that the 2CK model maps to a
Majorana resonant-level model \cite{emery1992,sengupta1994,ye1996};
one Majorana component remains decoupled from the rest of the system
and it leads to the fractional residual impurity entropy
\cite{emery1992,bolech2006}. Similar mechanism is at play in the
present model.

A quantum dot coupled to the TSC and ferromagnetic electrode will not,
in general, exhibit the full O(3) symmetry (as defined, for example,
in Ref.~\onlinecite{bradley1999}), thus one of the crucial questions
is whether the non-Fermi-liquid (NFL) fixed point exists under more
general conditions. The NRG calculations (described below) show that a
sufficient condition for obtaining the NFL state is that one of the
impurity Majorana modes is fully decoupled and remains uncompensated
at low temperatures: the asymptotic approach to the $T=0$ fixed point
is then always found to correspond to that in the two-channel Kondo
model. This is in line with the observation made in
Ref.~\onlinecite{bradley1999} which emphasizes the presence of the
zero mode which results in the singular scattering of the renormalized
Majorana fermions; the decoupled mode is important for the emergence
of the NFL state, not the O(3) symmetry on high energy scales. To tune
the system to the NFL fixed point, one may change the gate voltage and
apply an external magnetic field (similar procedure is applied in
quantum dots coupled to ferromagnetic leads, where tuning is necessary
to restore the Kondo effect, see
Ref.~\onlinecite{martinek2003a,martinek2003b,choi2004}). If the system
is not fully tuned to the NFL fixed point, but it is near it, there
will be a finite temperature range where the NFL behavior can be
observed, before the cross-over to the FL ground state
\cite{andrei1995}.

We study the resulting quantum impurity problem using the numerical
renormalization group (NRG) \cite{wilson1975, krishna1980a,
krishna1980b, bulla2008}. The method consists of discretizing the
continuum of the conduction band electrons, tridiagonalising the
resulting discrete Hamiltonian so that it takes the form of a
semi-infinite tight-binding chain with geometrically decreasing
hopping constants (Wilson chain), and diagonalizing this chain
Hamiltonian in an iterative fashion by taking into account one further
site in each renormalization-group transformation step. The
discretization is controlled by a parameter $\Lambda>1$, so that the
discretization intervals are $(\Lambda^{-(n+1)}:\Lambda^{-n})$; in
this work, $\Lambda=3$ in most calculations. The results are improved
by performing twist averaging with $N_z=4$ different discretization
meshes \cite{oliveira1994, campo2005, resolution}. The spectral
functions are computed using the density-matrix approach with complete
Fock space \cite{hofstetter2000, peters2006, weichselbaum2007}, and
the conductance curves at finite temperatures are obtained using the
Meir-Wingreen formula from the spectral data \cite{meir1992, costi2001,
yoshida2009prb}:
\begin{equation}
G_\downarrow(T) = \frac{e^2}{h} \pi \Gamma_\downarrow
\int_{-\infty}^{\infty} d\omega \left( - \frac{\partial f}{\partial
\omega} \right) A_\downarrow(\omega,T),
\end{equation}
where $f(\omega)=[1+\exp(\beta\omega)]^{-1}$ is the Fermi function,
$\beta=1/k_B T$ and the chemical potential has been fixed at zero
energy, while $A_\downarrow(\omega,T)$ is the spin-down spectral
function on the impurity site. Note that we are only considering the
linear conductance for the spin-down electrons which corresponds to
the spin-polarized transport flowing from the ferromagnetic source to
the ferromagnetic drain electrode. The spin-down electrons are
conserved and there is no mixing between the spin-up and spin-down
electrons (in the absence of the magnetic field in the transverse
direction, i.e., in the $x$-$y$ plane). In general, the Hamiltonian
has no symmetries which could be used to simplify the calculations by
the Wigner-Eckart theorem. It is thus necessary to diagonalize one
large matrix in each NRG step. It is important to keep enough states
in the NRG truncation to prevent spurious symmetry breaking. The NRG
implementation has been tested by performing calculations for a
non-interacting Majorana resonant-level model; see also
Appendix~\ref{app}. An excellent agreement is found between the
numerical and the exact analytical results.

\section{Results}
\label{sec3}

\subsection{Thermodynamics}

We first study the impurity contribution to the total electronic
entropy, defined as
\begin{equation}
S_\mathrm{imp}(T)=S(T)-S^{(0)}(T),
\end{equation}
where $S(T)$ is the entropy for the full problem while $S^{(0)}(T)$ is
the entropy for the problem without the impurity. Thus
$S_\mathrm{imp}(T)$ measures the effective degrees of freedom on the
impurity site on the temperature scale $T$. 
We use the parametrization
\begin{equation}
t_1=t \cos\alpha,\quad t_2=t \sin\alpha,
\end{equation}
where $\alpha=\pi/4$ corresponds to the regular Anderson impurity model
($A=0$) and $\alpha=0$ to the model with one fully decoupled Majorana
channel ($V=A$). The overall hybridization $t$ is chosen so that
$\Gamma_\uparrow=\Gamma_\downarrow$ in the $t_1=t_2$ limit.

\begin{figure}[htbp!]
\centering
\includegraphics[clip,width=7cm]{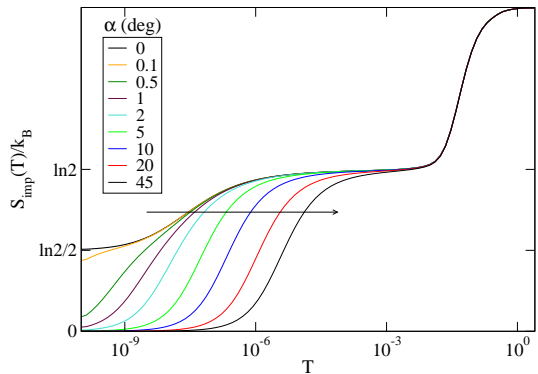}
\caption{(Color online) The impurity contribution to the electron
entropy as a function of the temperature for different values of the
parameter $\alpha$ which quantifies the ratio of Majorana hopping
rates. The model parameters are $U=0.2$,
$\Gamma_\uparrow=\Gamma_\downarrow=0.01$ (for the $t_1=t_2$ limit),
and $\epsilon+U/2=0$. The arrow indicates the direction of increasing
$\alpha$.}
\label{fig1}
\end{figure}

The resulting impurity entropy curves are shown in Fig.~\ref{fig1}. At
$T \sim U$ the system crosses over from the high-temperature
free-impurity fixed point (where the impurity level can be found in
either of the four states with equally probability, hence the $\ln 4$
impurity entropy) to the local-moment fixed point (where only the spin
can fluctuate, since the charge fluctuations are frozen, hence the
$\ln 2$ impurity entropy). For $\alpha=\pi/4=45\degg$, the system then
undergoes the conventional single-channel spin-$1/2$ Kondo effect at
$T \sim T_K$ in which the impurity spin degree of freedom is fully
screened; here $T_K$ is approximately given by the Haldane formula
\cite{haldane1978, krishna1980a, krishna1980b, hewson}
\begin{equation}
T_K = 0.182 U \sqrt{\rho J_K} \exp\left( - \frac{1}{\rho J_K} \right),
\end{equation}
with $\rho J_K=8\Gamma/\pi U$. This expression is valid for
$\epsilon+U/2=0$, i.e., when the system is at the particle-hole
symmetric point, and $\Gamma=\Gamma_\uparrow=\Gamma_\downarrow$. If
$\Gamma_\uparrow \neq \Gamma_\downarrow$, one has to use the theory
for the Kondo effect in the presence of the itinerant-electron
ferromagnetism \cite{martinek2003a, martinek2003b,choi2004}, the main
effect of which is the modification of the exponential factor to
\begin{equation}
\exp\left( - \frac{1}{\rho J_K} \frac{\mathrm{arctanh}P}{P} \right),
\end{equation}
where $P=(\Gamma_\uparrow - \Gamma_\downarrow) / (\Gamma_\uparrow +
\Gamma_\downarrow)$ is the spin polarization, thus the Kondo scale is
accordingly reduced. The results in Fig.~\ref{fig1} show that the
Kondo scale is also reduced if the ratio between the hopping
parameters for the two Majorana modes of spin-up electrons is detuned
from the symmetric $t_1=t_2$ case. For a wide range of parameters
$\alpha$, the entropy curves simply follows the universal
single-channel $S=1/2$ Kondo model entropy curve, the only effect is
the reduced Kondo temperature. In other words, the curves overlap if
shifted horizontally (on the logarithmic scale). Only for $\alpha <
4\degg$ can one observe different behavior: while the asymptotic tails
($T \ll T_K$) still follow the universal curve, the cross-over curves
($T \sim T_K$) exhibit slower temperature variation. For very small
$\alpha < 0.2 \degg$ one can observe a two-stage behavior: the system
first goes to a non-Fermi-liquid fixed point with $\ln2/2$ entropy,
but since this fixed point is unstable, there is another cross-over to
a final Fermi-liquid ground state at some lower temperature
\cite{andrei1995}. Only for exactly $\alpha=0$ is the NFL fixed point
stable and the system has residual entropy down to zero temperature.
As expected, the entropy curves can be fitted with the entropy curves
calculated for the two-channel Kondo model with channel asymmetry
($J_1 \neq J_2$); the channel-symmetric case ($J_1 = J_2$) corresponds
to the $\alpha=0$ limit of the present model \cite{bulla1997sigmatau}.

\begin{figure}[htbp!]
\centering
\includegraphics[clip,width=6cm]{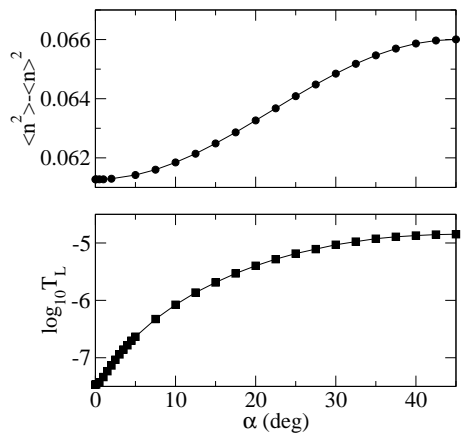}
\caption{a) Charge fluctuations as a function of the parameter
$\alpha$. b) The low-temperature scale of the problem, defined as
$S(T_L)=3\ln4/4$. In the regime where the regular Kondo effect scaling
is observed (roughly $\alpha>5\degg$), $T_L$ approximately corresponds
to the Kondo temperature $T_K$.}
\label{fig2}
\end{figure}

In Fig.~\ref{fig2} we show charge fluctuations and the low-temperature
scale $T_L$ of the problem as a function of $\alpha$ (for small
$\alpha$, there are actually two different low-temperature scales, one
associated with the Kondo screening and the other with the cross-over
from the NFL to the Fermi-liquid (FL) state; the results for $T_L$ are
actually meaningful only for large $\alpha>5\degg$, where they roughly
correspond to the Kondo temperature of the conventional Kondo
screening). We see that reducing $\alpha$ leads to a small reduction
of charge fluctuations (by approximately $7\%$); this corresponds to
the gradual freezing-out of the fluctuations of one of the Majorana
modes. The low-temperature scale of the problem decreases accordingly.
This behavior is similar to that found in the ferromagnetic Kondo
problem, where with the increasing polarization $P$ the charge
fluctuations of {\sl both} spin species are reduced [even though the
average hybridization $(\Gamma_\uparrow+\Gamma_\downarrow)/2$ remains
constant, thus the hybridization of one spin species decreases while
that of the other actually increases] and the Kondo temperature is
exponentially lowered.

\begin{figure}[htbp!]
\centering
\includegraphics[clip,width=8cm]{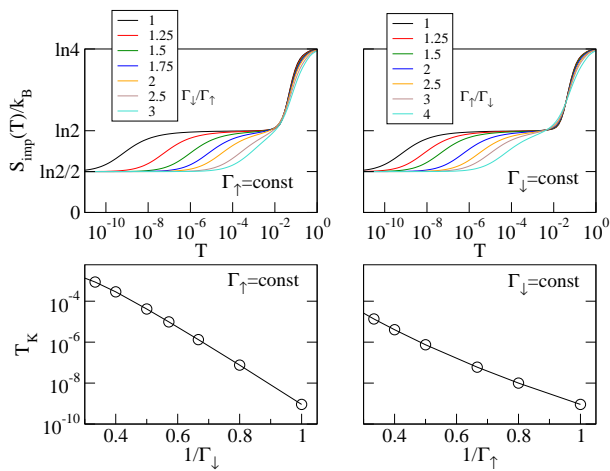}
\caption{(Color online) The temperature scales as a function
of the hybridization with the Majorana mode ($\Gamma_\uparrow$)
and the measurement leads ($\Gamma_\downarrow$). In each case
one of the hybridization parameters is held fixed at a value of
$\Gamma_\sigma=0.01$, while the other is varied. $U=0.2$, $\epsilon_d=-U/2$.}
\label{fign1}
\end{figure}

The Kondo temperature in the NFL regime ($t_2=0$) can be tuned by
changing either of the two hybridization parameters, $\Gamma_\uparrow$
(that is, the coupling to the TS) or $\Gamma_\downarrow$ (the coupling
to the probe leads). In both cases the dependence is exponential, see
Fig.~\ref{fign1}.  We reiterate in passing that $\Gamma_\uparrow$ and
$\Gamma_\downarrow$ by no means have to be equal for the two-channel
Kondo effect to emerge.

\subsection{Transport properties}

\begin{figure}[htbp!]
\centering
\includegraphics[clip,width=7cm]{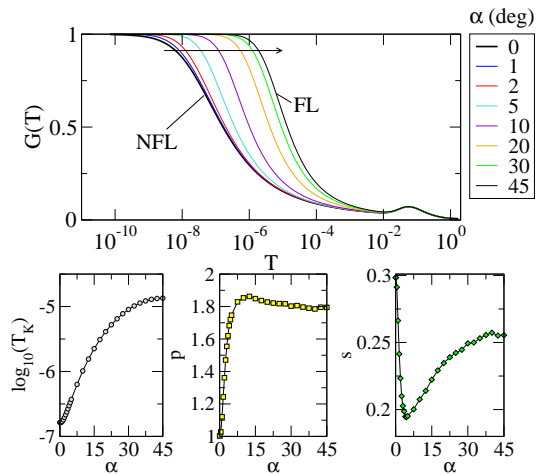}
\caption{(Color online) a) Temperature dependence of the conductance
through the quantum dot for different values of $\alpha$. The arrow
indicates the direction of increasing $\alpha$. b,c,d) The variation
of the fit parameters $T_K$, $p$, and $s$ as a function of $\alpha$. }
\label{fig3}
\end{figure}

In Fig.~\ref{fig3} we plot one of the main results of this work, the
temperature dependence of the linear conductance as measured between
the probe source and drain electrodes. The subfigures b,c,d
show the results of a fit using an empirical function
\cite{goldhabergordon1998a, parks2010}
\begin{equation}
\label{ansatz}
G(T) = G_0 \left[ 1 + (2^{1/s}-1) (T/T_K)^p \right]^{-s},
\end{equation}
where $T_K$ is defined as $G(T_K)=G(0)/2$, $p$ describes the exponent
of the asymptotic behavior for small $T$ (Fermi liquid behavior
corresponds to a $T^2$ finite-temperature correction, while for the
two-channel Kondo model NFL fixed-point one expects a linear
finite-temperature correction), while the parameter $s$ controls the
shape of the cross-over part of the curve. We find that the parameter
$T_K$ varies similarly as the low-temperature scale $T_L$ discussed
previously. The shape parameter $s$ is at first decreasing, but in the
low-$\alpha$ regime where the two-stage behavior starts to emerge
(roughly $\alpha<5\degg$) the conductance curves start to strongly
reflect the non-Fermi-liquid behavior at low temperature scales. This
is most strikingly visible in the behavior of the exponent parameter
$p$ which rapidly decreases towards the expected limiting value of
$p=1$. It is interesting to note that in the case of regular Anderson
impurity model (i.e., for $\alpha=45\degg$), the best fit is not
obtained for the standard values $p=2$, $s=0.22$, but rather for $p
\approx 1.8$, $s \approx 0.25$. This is due to the fact that the true
$T^2$ behavior only emerges asymptotically for $T \ll T_K$, where the
conductance is very close to the unitary limit, while in the
cross-over regime a better description is obtained with an effective
exponent different from 2. This is an important message for the
experimentalists: a deviation of the extracted parameter $p$ from the
value of 2 does not immediately imply non-Fermi-liquid properties of
the system at low temperatures, especially if the fit is performed in
the cross-over region. An extracted value approaching $p=1$ would,
however, constitute a ``smoking gun'' that the system is near the
two-channel Kondo model fixed point. Since the transport curves are
universal, the proposed transport experiment would thus consist of
measuring the conductance across one or two decades of temperatures
(around and below $T_K$, for example) and fitting with the $G(T)$
curves. It is not necessary to go to very low temperatures ($T \ll
T_K$) and try to extract the $T^2$ or $T$ scaling behavior; even on 
the scale of $T \sim T_K$ the universal $G(T)$ in both cases are
sufficiently different that one should be able to distinguish the two
situations (a comparison between the measured $G(T)$ curves and the
NRG calculations, for example, shows good agreement and has been used
to distinguish between the Kondo and the mixed-valence regimes
\cite{goldhabergordon1998a}, or between the Kondo effects with
different impurity spins \cite{parks2010}).

Finally, we must address the role of the gate voltage and the magnetic
field. Both types of operators are relevant (in the renormalization
group sense), since in the language of the Majorana fermions they
correspond to various coupling terms such as $d^\dag_\sigma
d_\sigma=-i \xi_{1\sigma} \xi_{2\sigma}$ where $\xi_{i\sigma}$ are the
Majorana modes of the impurity. Strictly speaking, the
non-Fermi-liquid fixed point is only stable at the particle-hole
symmetric point ($\epsilon+U/2=0$) and for zero external magnetic
field ($B=0$), thus the system needs to be tuned appropriately to
observe the two-channel Kondo effect. Note, however, that we have
assumed particle-hole symmetric flat bands. In general, the bands will
have some non-trivial density of states. In this case, the NFL fixed
point will be shifted away from the $\epsilon+U/2=0$, $B=0$ point and
the condition for observing the 2CK effect is such that the induced
magnetic and electric field in the quantum dot are compensated. This is
similar to the physics of the Kondo effect in the presence of
ferromagnetic leads \cite{martinek2003a,
martinek2003b,choi2004,pasupathy2004}.

It is worth noting that the two-channel Kondo effect may, in principle
at least, be easier to achieve in this system than in the
semiconductor quantum dot implementation of
Ref.~\onlinecite{potok2007}. In the latter system, the NFL fixed point
is achieved by using a larger (but interacting) quantum dot to
effectively play the role of the second channel (inter-channel
particle exchange is {\sl dynamically} prohibited by the penalty of
the charging energy); this then requires a subtle tuning to obtain
equal coupling to both channels, $J_1=J_2$. In the proposed system,
there are always only three Majorana channels, and one solely needs to
tune the quantum dot parameters such that one of the local Majorana
modes decouples. (In this respect the problem is similar to the case
of a QD coupled to the edge states of the FQHE \cite{fiete2008}, where
one also needs to tune solely the QD parameters. The required channel
symmetry is automatically present.)

\subsection{Magnetic field effects}

\begin{figure}[htbp!]
\centering
\includegraphics[clip,width=7cm]{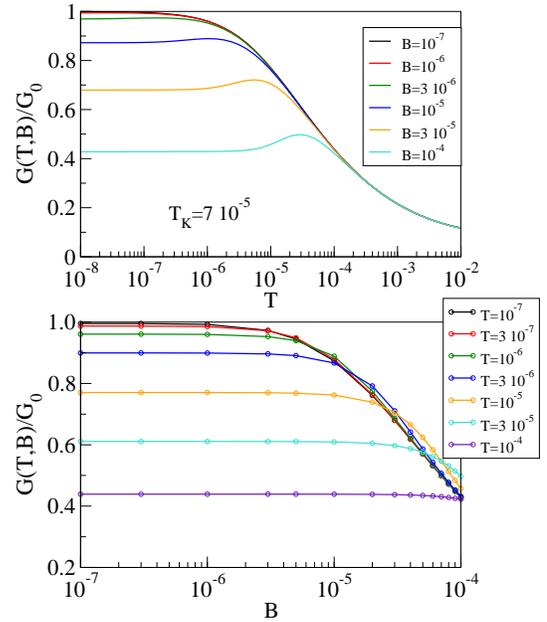}
\caption{(Color online) Temperature and magnetic-field dependence of
the conductance through the quantum dot. Model parameters are $U=0.2$,
$\epsilon=-U/2$, $\Gamma_\uparrow=0.03$, and $\Gamma_\downarrow=0.01$.}
\label{fign2}
\end{figure}

The system may also be probed at constant temperature by applying an
external magnetic field (which is assumed to couple only to the dot 
spin without perturbing other parts of the system). In the standard
Kondo effect, the magnetic field reduces the linear conductance at
$T=0$ quadratically for small $B$ \cite{costi2000, hewson}. In fact,
one may use a fitting function similar to Eq.~\eqref{ansatz}:
\begin{equation}
G(T=0,B) = G_0 \left[ 1+(2^{1/s'}-1)(B/T_K)^{p'} \right]^{-s'},
\end{equation}
where $B$ is expressed in temperature units $(g\mu_B/k_B$). When a fit
is performed for a FL regime over an interval of magnetic fields from
$B=0$ to $B=T_K$, one obtains for the exponent $p'=2$ and for the
shape parameter $s'=0.5$. Performing the same calculation for our
system in the NFL regime, we obtain, instead, the exponent $p'=1.3$
and the shape parameter $s'=0.36$. More extensive set of results for
the conductance at finite temperature and magnetic field are shown in
Fig.~\ref{fign2}. The results in the FL and NFL regimes are
characteristically different and allow for an additional measurement
approach.

\section{Conclusion}

It was shown that if a quantum impurity described as a single
interacting level is coupled to three independent Majorana channels,
but is decoupled from the fourth, a non-Fermi-liquid state emerges
which can be probed by performing linear conductance measurements. By
tuning the system parameters (in particular the gate voltage) the
non-Fermi-liquid regime can be obtained even in situations which do
not have the full O(3) symmetry between the three Majorana channels.
The experimental realization of the predicted effect could make use of
two QAH systems to provide fully spin-polarized complex fermions, and
one TSC system to provide the Majorana fermions of the opposite spin.
The experimental challenge thus consists -- in the first place -- in
actually creating the QAH and TSC systems, and in establishing the
electrical contacts between the quantum dot and these systems. The
non-Fermi-liquid state should then naturally emerge and it should be
rather robust (as robust as the edge states themselves). Further
complications might arise from the interactions between the electrons
in the one-dimension channels, since they might drive the system to a
different fixed point.

\appendix

\section{Majorana resonant-level model}
\label{app}

For reference purposes (and for testing the numerical method) we now
solve exactly the resonant-level model with different couplings to
the two Majorana modes of a single-channel continuum. The spin
index plays no role in a non-interacting model, thus we omit it in
writing. The Hamiltonian is composed of the following terms:
\begin{equation}
\begin{split}
H_0 &= \epsilon n_\sigma, \\
H_1 &= \sum_{k} \epsilon_k c^\dag_{k}c_{k},\\
H_2 &= \sum_{k} \left( V c^\dag_k d + A c^\dag_k d^\dag + \text{H.c.}
\right).
\end{split}
\end{equation}
We assume that the hopping coefficients $V$ and $A$ do not depend on
$k$, and for simplicity we take them to be real. We will use the
notation $\kor{A;B}_z$ for a correlator between the operators $A$
and $B$, and at the end the argument $z$ will be chosen as
$z=\omega+i\delta$ to obtain the retarded Green's functions
($\delta\to0$). We are particularly interested in the Green's function
$G(\omega)=\kor{d;d^\dag}_{\omega+i\delta}$ which gives the spectral
function as $A(\omega)=(-1/\pi) \Imm G(\omega)$. We use the equation of
motion method:
\begin{equation}
z \kor{A;B}_z = \expv{[A,B]_\eta} - \kor{A;[B,H]_-}_z,
\end{equation}
where $\eta=+$ (anticommutator) if $A$ and $B$ are both fermionic
operators, and $\eta=-$ (commutator) in all other cases.

We introduce the notation $g=\kor{d;d^\dag}_z$ and $h=\kor{d;d}$, as
well as $g_k=\kor{d;c_{k}^\dag}$ and $h_k=\kor{d;c_k}$. The equations
of motion then give
\begin{equation}
\label{eqs}
\begin{split}
(z-\epsilon)g&=1+V \sum_k g_k - A \sum_k h_k, \\
(z-\epsilon_k)g_k&=V g+A h, \\
(z+\epsilon)h&=-V \sum_k h_k+A \sum_k g_k, \\
(z+\epsilon_k)h_k&=-A g-V h.
\end{split}
\end{equation}
We introduce $\gamma_1(z)=\sum_k 1/(z-\epsilon_k)$ and
$\gamma_2(z)=\sum_k 1/(z+\epsilon_k)$, whose imaginary parts for
argument $z=\omega+i\delta$ are proportional to the density of states
in the lead, $\rho(\omega)$. For a particle-hole symmetric band,
$\gamma_1$ and $\gamma_2$ are fully equivalent. Expressing $g_k$ and
$h_k$ in terms of $g$ and $h$, inserting them in the equations of
motion for $g$ and $h$, then solving the resulting equations for $g$,
we obtain
\begin{equation}
g=\frac{\epsilon+z-(A^2 \gamma_1+V^2 \gamma_2)}
{(c\gamma_1-\epsilon)(c\gamma_2+\epsilon)-d(\gamma_1+\gamma_2)z
+z^2},
\end{equation}
where $c=A^2-V^2$ and $d=A^2+V^2$. In the wide-band limit,
$\gamma_{1,2} \to -i \pi \rho$, where $\rho$ is the constant density
of states. The spectral function is then
\begin{widetext}
\begin{equation}
A(\omega) \approx \frac{1}{\pi}
\frac{\pi \rho d \left[(\pi \rho c)^2-(\omega+\epsilon)^2\right]}
{[(\pi\rho c)^2+\epsilon^2]^2+2[(\pi \rho)^2 (A^4+6 A^2
V^2+V^4)-\epsilon^2]\omega^2+\omega^4}.
\end{equation}

In the particle-hole symmetric case ($\epsilon=0$), the half-width at
half-maximum of the spectral function is
\begin{equation}
\Gamma=\pi\rho \left[ \left(A^8-4 A^6 V^2+70 A^4 V^4-4 A^2
V^6+V^8\right)^{1/2}-8 A^2V^2 \right]^{1/2}.
\end{equation}
\end{widetext}
For $A=0$, this expression reduces to the expected result
$\Gamma=\pi\rho V^2$. In the $V=0$ limit, the result is $\Gamma=\pi
\rho A^2$. For $|A| \to |V|$ the width of the resonance goes to zero
and a delta peak emerges in the spectral function at $\omega=0$, see
also Ref.~\onlinecite{bradley1999}. This corresponds to the case of a
fully decoupled Majorana mode. Strictly speaking, the system is then
in a NFL state with $\ln 2/2$ residual entropy (per spin). The delta
peak carries half the spectral weight and there is a broader
background peak associated with the hybridized Majorana partner of the
decoupled mode; this broad spectral peak carries the remaining half of
the spectral weight.

If the problem is not particle-hole symmetric ($\epsilon \neq 0$) the
two Majorana modes remain coupled through the charge term (since
$d^\dag d = -i\eta_1 \eta_2$). In this case there can be no decoupled
Majorana mode and at zero temperature the system is in a FL state for
all values of $A$ and $V$. As $|A| \to |V|$, the spectral function
will have a maximum at $\omega \approx \epsilon$ (with a shift of the
order of the spectral peak width $\Gamma$) and will touch zero exactly
at $\omega=-\epsilon$.

The anomalous Green's function $h(z)=\kor{d;d}_z$ is
\begin{equation}
h=\frac{A V(\gamma_1+\gamma_2)}
{(c\gamma_1-\epsilon)(c\gamma_2-\epsilon)-d(\gamma_1+\gamma_2)z+z^2}.
\end{equation}
It is proportional to $A$, thus it vanishes in the absence of the
anomalous hybridization. In the wide-band limit, the anomalous
spectral function $B(\omega)=(-1/\pi)\Imm h(\omega+i\delta)$ is
\begin{widetext}
\begin{equation}
B(\omega) \approx \frac{-1}{\pi} \frac{2\pi\rho AV
\left( (\pi\rho c)^2 + (\epsilon-\omega)(\epsilon+\omega) \right)}
{[(\pi\rho c)^2+\epsilon^2]^2+2[(\pi \rho)^2 (A^4+6 A^2
V^2+V^4)-\epsilon^2]\omega^2+\omega^4}.
\end{equation}
\end{widetext}

In the particle-hole symmetric case ($\epsilon=0$) this spectral
function has an inverted (negative) peak at $\omega=0$ superimposed on
a broader positive resonance. In the $|A| \to |V|$ limit, the inverted
peak narrows down until it becomes a delta peak. This feature thus
corresponds to the decoupled Majorana mode, while the positive broad
resonance corresponds to its Majorana partner state.

For $\epsilon \neq 0$, the spectral function $B(\omega)$ goes through
zero always at $\omega=\pm\epsilon$, i.e., $|\epsilon|$ sets the scale
of the inverted spectral peak. As $|A| \to |V|$ only the weight of
this peak saturates, while the width remains roughly constant, since
the Majorana mode does not decouple.

\begin{acknowledgments}
R.Z. acknowledges the support of the Slovenian Research Agency (ARRS)
under Grant No. Z1-2058.
\end{acknowledgments}

\bibliography{majorana}

\begin{thebibliography}{80}
\expandafter\ifx\csname natexlab\endcsname\relax\def\natexlab#1{#1}\fi
\expandafter\ifx\csname bibnamefont\endcsname\relax
  \def\bibnamefont#1{#1}\fi
\expandafter\ifx\csname bibfnamefont\endcsname\relax
  \def\bibfnamefont#1{#1}\fi
\expandafter\ifx\csname citenamefont\endcsname\relax
  \def\citenamefont#1{#1}\fi
\expandafter\ifx\csname url\endcsname\relax
  \def\url#1{\texttt{#1}}\fi
\expandafter\ifx\csname urlprefix\endcsname\relax\def\urlprefix{URL }\fi
\providecommand{\bibinfo}[2]{#2}
\providecommand{\eprint}[2][]{\url{#2}}

\bibitem[{\citenamefont{von Klitzing et~al.}(1980)\citenamefont{von Klitzing,
  Dorda, and Pepper}}]{klitzing1980}
\bibinfo{author}{\bibfnamefont{K.}~\bibnamefont{von Klitzing}},
  \bibinfo{author}{\bibfnamefont{G.}~\bibnamefont{Dorda}}, \bibnamefont{and}
  \bibinfo{author}{\bibfnamefont{M.}~\bibnamefont{Pepper}},
  \bibinfo{journal}{Phys. Rev. Lett.} \textbf{\bibinfo{volume}{45}},
  \bibinfo{pages}{494} (\bibinfo{year}{1980}).

\bibitem[{\citenamefont{Thouless et~al.}(1982)\citenamefont{Thouless, Kohmoto,
  Nightingale, and den Nijs}}]{thouless1982}
\bibinfo{author}{\bibfnamefont{D.~J.} \bibnamefont{Thouless}},
  \bibinfo{author}{\bibfnamefont{M.}~\bibnamefont{Kohmoto}},
  \bibinfo{author}{\bibfnamefont{M.~P.} \bibnamefont{Nightingale}},
  \bibnamefont{and} \bibinfo{author}{\bibfnamefont{M.}~\bibnamefont{den Nijs}},
  \bibinfo{journal}{Phys. Rev. Lett.} \textbf{\bibinfo{volume}{49}},
  \bibinfo{pages}{405} (\bibinfo{year}{1982}).

\bibitem[{\citenamefont{Kane and Mele}(2005)}]{kane2005f}
\bibinfo{author}{\bibfnamefont{C.~L.} \bibnamefont{Kane}} \bibnamefont{and}
  \bibinfo{author}{\bibfnamefont{E.~J.} \bibnamefont{Mele}},
  \bibinfo{journal}{Phys. Rev. Lett.} \textbf{\bibinfo{volume}{95}},
  \bibinfo{pages}{146802} (\bibinfo{year}{2005}).

\bibitem[{\citenamefont{Fu and Kane}(2006)}]{fu2006f}
\bibinfo{author}{\bibfnamefont{L.}~\bibnamefont{Fu}} \bibnamefont{and}
  \bibinfo{author}{\bibfnamefont{C.~L.} \bibnamefont{Kane}},
  \bibinfo{journal}{Phys. Rev. B} \textbf{\bibinfo{volume}{74}},
  \bibinfo{pages}{195312} (\bibinfo{year}{2006}).

\bibitem[{\citenamefont{Bernevig and Zhang}(2006)}]{bernevig2006f}
\bibinfo{author}{\bibfnamefont{B.~A.} \bibnamefont{Bernevig}} \bibnamefont{and}
  \bibinfo{author}{\bibfnamefont{S.~C.} \bibnamefont{Zhang}},
  \bibinfo{journal}{Phys. Rev. Lett.} \textbf{\bibinfo{volume}{96}},
  \bibinfo{pages}{106802} (\bibinfo{year}{2006}).

\bibitem[{\citenamefont{Bernevig et~al.}(2006)\citenamefont{Bernevig, Hughes,
  and Zhang}}]{bernevig2006sc}
\bibinfo{author}{\bibfnamefont{B.~A.} \bibnamefont{Bernevig}},
  \bibinfo{author}{\bibfnamefont{T.~L.} \bibnamefont{Hughes}},
  \bibnamefont{and} \bibinfo{author}{\bibfnamefont{S.~C.} \bibnamefont{Zhang}},
  \bibinfo{journal}{Science} \textbf{\bibinfo{volume}{314}},
  \bibinfo{pages}{1757} (\bibinfo{year}{2006}).

\bibitem[{\citenamefont{Moore and Balents}(2007)}]{moore2007f}
\bibinfo{author}{\bibfnamefont{J.~E.} \bibnamefont{Moore}} \bibnamefont{and}
  \bibinfo{author}{\bibfnamefont{L.}~\bibnamefont{Balents}},
  \bibinfo{journal}{Phys. Rev. B} \textbf{\bibinfo{volume}{75}},
  \bibinfo{pages}{121306(R)} (\bibinfo{year}{2007}).

\bibitem[{\citenamefont{K\"onig et~al.}(2007)}]{konig2007d}
\bibinfo{author}{\bibfnamefont{M.}~\bibnamefont{K\"onig}} \bibnamefont{et~al.},
  \bibinfo{journal}{Science} \textbf{\bibinfo{volume}{318}},
  \bibinfo{pages}{766} (\bibinfo{year}{2007}).

\bibitem[{\citenamefont{Bergman and Hur}(2009)}]{bergman2009}
\bibinfo{author}{\bibfnamefont{D.~L.} \bibnamefont{Bergman}} \bibnamefont{and}
  \bibinfo{author}{\bibfnamefont{K.~L.} \bibnamefont{Hur}},
  \bibinfo{journal}{Phys. Rev. B} \textbf{\bibinfo{volume}{79}},
  \bibinfo{pages}{184520} (\bibinfo{year}{2009}).

\bibitem[{\citenamefont{Nilsson et~al.}(2008)\citenamefont{Nilsson, Akhmerov,
  and Beenakker}}]{nilsson2008}
\bibinfo{author}{\bibfnamefont{J.}~\bibnamefont{Nilsson}},
  \bibinfo{author}{\bibfnamefont{A.~R.} \bibnamefont{Akhmerov}},
  \bibnamefont{and} \bibinfo{author}{\bibfnamefont{C.~W.~J.}
  \bibnamefont{Beenakker}}, \bibinfo{journal}{Phys. Rev. Lett.}
  \textbf{\bibinfo{volume}{101}}, \bibinfo{pages}{120403}
  (\bibinfo{year}{2008}).

\bibitem[{\citenamefont{Ivanov}(2001)}]{ivanov2001}
\bibinfo{author}{\bibfnamefont{D.~A.} \bibnamefont{Ivanov}},
  \bibinfo{journal}{Phys. Rev. Lett.} \textbf{\bibinfo{volume}{86}},
  \bibinfo{pages}{268} (\bibinfo{year}{2001}).

\bibitem[{\citenamefont{Stern}(2010)}]{stern2010}
\bibinfo{author}{\bibfnamefont{A.}~\bibnamefont{Stern}},
  \bibinfo{journal}{Nature} \textbf{\bibinfo{volume}{464}},
  \bibinfo{pages}{187} (\bibinfo{year}{2010}).

\bibitem[{\citenamefont{Kitaev}(2001)}]{kitaev2001}
\bibinfo{author}{\bibfnamefont{A.}~\bibnamefont{Kitaev}},
  \bibinfo{journal}{Usp. Fiz. Nauk (Suppl.)} \textbf{\bibinfo{volume}{171}},
  \bibinfo{pages}{131} (\bibinfo{year}{2001}).

\bibitem[{\citenamefont{Kitaev}(2003)}]{kitaev2003}
\bibinfo{author}{\bibfnamefont{A.}~\bibnamefont{Kitaev}},
  \bibinfo{journal}{Ann. Phys.} \textbf{\bibinfo{volume}{303}},
  \bibinfo{pages}{2} (\bibinfo{year}{2003}).

\bibitem[{\citenamefont{Wilczek}(2009)}]{wilczek2009}
\bibinfo{author}{\bibfnamefont{F.}~\bibnamefont{Wilczek}},
  \bibinfo{journal}{Nat. Phys.} \textbf{\bibinfo{volume}{5}},
  \bibinfo{pages}{614} (\bibinfo{year}{2009}).

\bibitem[{\citenamefont{Fu and Kane}(2008)}]{fu2008majorana}
\bibinfo{author}{\bibfnamefont{L.}~\bibnamefont{Fu}} \bibnamefont{and}
  \bibinfo{author}{\bibfnamefont{C.~L.} \bibnamefont{Kane}},
  \bibinfo{journal}{Phys. Rev. Lett.} \textbf{\bibinfo{volume}{100}},
  \bibinfo{pages}{096407} (\bibinfo{year}{2008}).

\bibitem[{\citenamefont{Fu and Kane}(2009{\natexlab{a}})}]{fu2009inter}
\bibinfo{author}{\bibfnamefont{L.}~\bibnamefont{Fu}} \bibnamefont{and}
  \bibinfo{author}{\bibfnamefont{C.~L.} \bibnamefont{Kane}},
  \bibinfo{journal}{Phys. Rev. Lett.} \textbf{\bibinfo{volume}{102}},
  \bibinfo{pages}{216403} (\bibinfo{year}{2009}{\natexlab{a}}).

\bibitem[{\citenamefont{Sau et~al.}(2010{\natexlab{a}})\citenamefont{Sau,
  Lutchyn, Tewari, and Sarma}}]{sau2010}
\bibinfo{author}{\bibfnamefont{J.~D.} \bibnamefont{Sau}},
  \bibinfo{author}{\bibfnamefont{R.~M.} \bibnamefont{Lutchyn}},
  \bibinfo{author}{\bibfnamefont{S.}~\bibnamefont{Tewari}}, \bibnamefont{and}
  \bibinfo{author}{\bibfnamefont{S.~D.} \bibnamefont{Sarma}},
  \bibinfo{journal}{Phys. Rev. Lett.} \textbf{\bibinfo{volume}{104}},
  \bibinfo{pages}{040502} (\bibinfo{year}{2010}{\natexlab{a}}).

\bibitem[{\citenamefont{Sau et~al.}(2010{\natexlab{b}})\citenamefont{Sau,
  Tewari, Lutchyn, Stanescu, and Sarma}}]{sau2010prb}
\bibinfo{author}{\bibfnamefont{J.~D.} \bibnamefont{Sau}},
  \bibinfo{author}{\bibfnamefont{S.}~\bibnamefont{Tewari}},
  \bibinfo{author}{\bibfnamefont{R.~M.} \bibnamefont{Lutchyn}},
  \bibinfo{author}{\bibfnamefont{T.~D.} \bibnamefont{Stanescu}},
  \bibnamefont{and} \bibinfo{author}{\bibfnamefont{S.~D.} \bibnamefont{Sarma}},
  \bibinfo{journal}{Phys. Rev. B} \textbf{\bibinfo{volume}{82}},
  \bibinfo{pages}{214509} (\bibinfo{year}{2010}{\natexlab{b}}).

\bibitem[{\citenamefont{Fu and Kane}(2009{\natexlab{b}})}]{fu2009josephson}
\bibinfo{author}{\bibfnamefont{L.}~\bibnamefont{Fu}} \bibnamefont{and}
  \bibinfo{author}{\bibfnamefont{C.~L.} \bibnamefont{Kane}},
  \bibinfo{journal}{Phys. Rev. B} \textbf{\bibinfo{volume}{79}},
  \bibinfo{pages}{161408(R)} (\bibinfo{year}{2009}{\natexlab{b}}).

\bibitem[{\citenamefont{Tanaka et~al.}(2009)\citenamefont{Tanaka, Yokoyama, and
  Nagaosa}}]{tanaka2009nfis}
\bibinfo{author}{\bibfnamefont{Y.}~\bibnamefont{Tanaka}},
  \bibinfo{author}{\bibfnamefont{T.}~\bibnamefont{Yokoyama}}, \bibnamefont{and}
  \bibinfo{author}{\bibfnamefont{N.}~\bibnamefont{Nagaosa}},
  \bibinfo{journal}{Phys. Rev. Lett.} \textbf{\bibinfo{volume}{103}},
  \bibinfo{pages}{107002} (\bibinfo{year}{2009}).

\bibitem[{\citenamefont{Akhmerov et~al.}(2009)\citenamefont{Akhmerov, Nilsson,
  and Beenakker}}]{akhmerov2009inter}
\bibinfo{author}{\bibfnamefont{A.~R.} \bibnamefont{Akhmerov}},
  \bibinfo{author}{\bibfnamefont{J.}~\bibnamefont{Nilsson}}, \bibnamefont{and}
  \bibinfo{author}{\bibfnamefont{C.~W.~J.} \bibnamefont{Beenakker}},
  \bibinfo{journal}{Phys. Rev. Lett.} \textbf{\bibinfo{volume}{102}},
  \bibinfo{pages}{216404} (\bibinfo{year}{2009}).

\bibitem[{\citenamefont{Law et~al.}(2009)\citenamefont{Law, Lee, and
  Ng}}]{law2009}
\bibinfo{author}{\bibfnamefont{K.~T.} \bibnamefont{Law}},
  \bibinfo{author}{\bibfnamefont{P.~A.} \bibnamefont{Lee}}, \bibnamefont{and}
  \bibinfo{author}{\bibfnamefont{T.~K.} \bibnamefont{Ng}},
  \bibinfo{journal}{Phys. Rev. Lett.} \textbf{\bibinfo{volume}{103}},
  \bibinfo{pages}{237001} (\bibinfo{year}{2009}).

\bibitem[{\citenamefont{Sau et~al.}(2010{\natexlab{c}})\citenamefont{Sau,
  Tewari, and Sarma}}]{sau2010pp}
\bibinfo{author}{\bibfnamefont{J.~D.} \bibnamefont{Sau}},
  \bibinfo{author}{\bibfnamefont{S.}~\bibnamefont{Tewari}}, \bibnamefont{and}
  \bibinfo{author}{\bibfnamefont{S.~D.} \bibnamefont{Sarma}},
  \emph{\bibinfo{title}{Probing non-abelian statistics with majorana fermion
  interferometry in spin-orbit-coupled semiconductors}},
  \bibinfo{howpublished}{arxiv:1004.4702} (\bibinfo{year}{2010}{\natexlab{c}}).

\bibitem[{\citenamefont{Fu}(2010)}]{fu2010tele}
\bibinfo{author}{\bibfnamefont{L.}~\bibnamefont{Fu}}, \bibinfo{journal}{Phys.
  Rev. Lett.} \textbf{\bibinfo{volume}{104}}, \bibinfo{pages}{056402}
  (\bibinfo{year}{2010}).

\bibitem[{\citenamefont{Hassler et~al.}(2010)\citenamefont{Hassler, Akhmerov,
  Hou, and Beenakker}}]{hassler2010}
\bibinfo{author}{\bibfnamefont{F.}~\bibnamefont{Hassler}},
  \bibinfo{author}{\bibfnamefont{A.~R.} \bibnamefont{Akhmerov}},
  \bibinfo{author}{\bibfnamefont{C.-Y.} \bibnamefont{Hou}}, \bibnamefont{and}
  \bibinfo{author}{\bibfnamefont{C.~W.~J.} \bibnamefont{Beenakker}},
  \bibinfo{journal}{New J. Phys.} \textbf{\bibinfo{volume}{12}},
  \bibinfo{pages}{125002} (\bibinfo{year}{2010}).

\bibitem[{\citenamefont{Fiete et~al.}(2008)\citenamefont{Fiete, Bishara, and
  Nayak}}]{fiete2008}
\bibinfo{author}{\bibfnamefont{G.~A.} \bibnamefont{Fiete}},
  \bibinfo{author}{\bibfnamefont{W.}~\bibnamefont{Bishara}}, \bibnamefont{and}
  \bibinfo{author}{\bibfnamefont{C.}~\bibnamefont{Nayak}},
  \bibinfo{journal}{Phys. Rev. Lett.} \textbf{\bibinfo{volume}{101}},
  \bibinfo{pages}{176801} (\bibinfo{year}{2008}).

\bibitem[{\citenamefont{Fiete et~al.}(2010)\citenamefont{Fiete, Bishara, and
  Nayak}}]{fiete2010}
\bibinfo{author}{\bibfnamefont{G.~A.} \bibnamefont{Fiete}},
  \bibinfo{author}{\bibfnamefont{W.}~\bibnamefont{Bishara}}, \bibnamefont{and}
  \bibinfo{author}{\bibfnamefont{C.}~\bibnamefont{Nayak}},
  \bibinfo{journal}{Phys. Rev. B} \textbf{\bibinfo{volume}{82}},
  \bibinfo{pages}{035301} (\bibinfo{year}{2010}).

\bibitem[{\citenamefont{Sevier and Fiete}(2011)}]{sevier2011}
\bibinfo{author}{\bibfnamefont{S.~A.} \bibnamefont{Sevier}} \bibnamefont{and}
  \bibinfo{author}{\bibfnamefont{G.~A.} \bibnamefont{Fiete}},
  \emph{\bibinfo{title}{Non-fermi liquid quantum impurity physics from
  non-abelian quantum hall states}}, \bibinfo{howpublished}{arxiv:1101.1326}
  (\bibinfo{year}{2011}).

\bibitem[{\citenamefont{Maldacena and Ludwig}(1997)}]{maldacena1997}
\bibinfo{author}{\bibfnamefont{J.~M.} \bibnamefont{Maldacena}}
  \bibnamefont{and} \bibinfo{author}{\bibfnamefont{A.~W.~W.}
  \bibnamefont{Ludwig}}, \bibinfo{journal}{Nucl. Phys. B}
  \textbf{\bibinfo{volume}{506}}, \bibinfo{pages}{565} (\bibinfo{year}{1997}).

\bibitem[{\citenamefont{Bulla et~al.}(1997)\citenamefont{Bulla, Hewson, and
  Zhang}}]{bulla1997sigmatau}
\bibinfo{author}{\bibfnamefont{R.}~\bibnamefont{Bulla}},
  \bibinfo{author}{\bibfnamefont{A.~C.} \bibnamefont{Hewson}},
  \bibnamefont{and} \bibinfo{author}{\bibfnamefont{G.-M.} \bibnamefont{Zhang}},
  \bibinfo{journal}{Phys. Rev. B} \textbf{\bibinfo{volume}{56}},
  \bibinfo{pages}{11721} (\bibinfo{year}{1997}).

\bibitem[{\citenamefont{Nozi{\`e}res and Blandin}(1980)}]{nozieres1980}
\bibinfo{author}{\bibfnamefont{P.}~\bibnamefont{Nozi{\`e}res}}
  \bibnamefont{and} \bibinfo{author}{\bibfnamefont{A.}~\bibnamefont{Blandin}},
  \bibinfo{journal}{J. Physique} \textbf{\bibinfo{volume}{41}},
  \bibinfo{pages}{193} (\bibinfo{year}{1980}).

\bibitem[{\citenamefont{Sacramento and Schlottmann}(1991)}]{sacramento1991}
\bibinfo{author}{\bibfnamefont{P.~D.} \bibnamefont{Sacramento}}
  \bibnamefont{and}
  \bibinfo{author}{\bibfnamefont{P.}~\bibnamefont{Schlottmann}},
  \bibinfo{journal}{Phys. Rev. B} \textbf{\bibinfo{volume}{43}},
  \bibinfo{pages}{13294} (\bibinfo{year}{1991}).

\bibitem[{\citenamefont{Affleck and Ludwig}(1992)}]{affleck1992}
\bibinfo{author}{\bibfnamefont{I.}~\bibnamefont{Affleck}} \bibnamefont{and}
  \bibinfo{author}{\bibfnamefont{A.~W.~W.} \bibnamefont{Ludwig}},
  \bibinfo{journal}{Phys. Rev. Lett.} \textbf{\bibinfo{volume}{68}},
  \bibinfo{pages}{1046} (\bibinfo{year}{1992}).

\bibitem[{\citenamefont{Emery and Kivelson}(1992)}]{emery1992}
\bibinfo{author}{\bibfnamefont{V.~J.} \bibnamefont{Emery}} \bibnamefont{and}
  \bibinfo{author}{\bibfnamefont{S.}~\bibnamefont{Kivelson}},
  \bibinfo{journal}{Phys. Rev. B} \textbf{\bibinfo{volume}{46}},
  \bibinfo{pages}{10812} (\bibinfo{year}{1992}).

\bibitem[{\citenamefont{Sengupta and Georges}(1994)}]{sengupta1994}
\bibinfo{author}{\bibfnamefont{A.~M.} \bibnamefont{Sengupta}} \bibnamefont{and}
  \bibinfo{author}{\bibfnamefont{A.}~\bibnamefont{Georges}},
  \bibinfo{journal}{Phys. Rev. B} \textbf{\bibinfo{volume}{49}},
  \bibinfo{pages}{10020(R)} (\bibinfo{year}{1994}).

\bibitem[{\citenamefont{Ye}(1996)}]{ye1996}
\bibinfo{author}{\bibfnamefont{J.}~\bibnamefont{Ye}}, \bibinfo{journal}{Phys.
  Rev. Lett.} \textbf{\bibinfo{volume}{77}}, \bibinfo{pages}{3224}
  (\bibinfo{year}{1996}).

\bibitem[{\citenamefont{Cox and Zawadowski}(1998)}]{cox1998}
\bibinfo{author}{\bibfnamefont{D.~L.} \bibnamefont{Cox}} \bibnamefont{and}
  \bibinfo{author}{\bibfnamefont{A.}~\bibnamefont{Zawadowski}},
  \bibinfo{journal}{Adv. Phys.} \textbf{\bibinfo{volume}{47}},
  \bibinfo{pages}{599} (\bibinfo{year}{1998}).

\bibitem[{\citenamefont{Zar{\'a}nd and von Delft}(2000)}]{zarand2000}
\bibinfo{author}{\bibfnamefont{G.}~\bibnamefont{Zar{\'a}nd}} \bibnamefont{and}
  \bibinfo{author}{\bibfnamefont{J.}~\bibnamefont{von Delft}},
  \bibinfo{journal}{Phys. Rev. B} \textbf{\bibinfo{volume}{61}},
  \bibinfo{pages}{6918} (\bibinfo{year}{2000}).

\bibitem[{\citenamefont{Zar\'and et~al.}(2002)\citenamefont{Zar\'and, Costi,
  Jerez, and Andrei}}]{zarand2002}
\bibinfo{author}{\bibfnamefont{G.}~\bibnamefont{Zar\'and}},
  \bibinfo{author}{\bibfnamefont{T.}~\bibnamefont{Costi}},
  \bibinfo{author}{\bibfnamefont{A.}~\bibnamefont{Jerez}}, \bibnamefont{and}
  \bibinfo{author}{\bibfnamefont{N.}~\bibnamefont{Andrei}},
  \bibinfo{journal}{Phys. Rev. B} \textbf{\bibinfo{volume}{65}},
  \bibinfo{pages}{134416} (\bibinfo{year}{2002}).

\bibitem[{\citenamefont{Pang and Cox}(1991)}]{pang1991}
\bibinfo{author}{\bibfnamefont{H.~B.} \bibnamefont{Pang}} \bibnamefont{and}
  \bibinfo{author}{\bibfnamefont{D.~L.} \bibnamefont{Cox}},
  \bibinfo{journal}{Phys. Rev. B} \textbf{\bibinfo{volume}{44}},
  \bibinfo{pages}{9454} (\bibinfo{year}{1991}).

\bibitem[{\citenamefont{Andrei and Jerez}(1995)}]{andrei1995}
\bibinfo{author}{\bibfnamefont{N.}~\bibnamefont{Andrei}} \bibnamefont{and}
  \bibinfo{author}{\bibfnamefont{A.}~\bibnamefont{Jerez}},
  \bibinfo{journal}{Phys. Rev. Lett.} \textbf{\bibinfo{volume}{74}},
  \bibinfo{pages}{4507} (\bibinfo{year}{1995}).

\bibitem[{\citenamefont{Potok et~al.}(2007)\citenamefont{Potok, Rau, Shtrikman,
  Oreg, and Goldhaber-Gordon}}]{potok2007}
\bibinfo{author}{\bibfnamefont{R.~M.} \bibnamefont{Potok}},
  \bibinfo{author}{\bibfnamefont{I.~G.} \bibnamefont{Rau}},
  \bibinfo{author}{\bibfnamefont{H.}~\bibnamefont{Shtrikman}},
  \bibinfo{author}{\bibfnamefont{Y.}~\bibnamefont{Oreg}}, \bibnamefont{and}
  \bibinfo{author}{\bibfnamefont{D.}~\bibnamefont{Goldhaber-Gordon}},
  \bibinfo{journal}{Nature} \textbf{\bibinfo{volume}{446}},
  \bibinfo{pages}{167} (\bibinfo{year}{2007}).

\bibitem[{\citenamefont{\v{Z}itko and Bon\v{c}a}(2007)}]{flnfl3}
\bibinfo{author}{\bibfnamefont{R.}~\bibnamefont{\v{Z}itko}} \bibnamefont{and}
  \bibinfo{author}{\bibfnamefont{J.}~\bibnamefont{Bon\v{c}a}},
  \bibinfo{journal}{Phys. Rev. Lett.} \textbf{\bibinfo{volume}{98}},
  \bibinfo{pages}{047203} (\bibinfo{year}{2007}).

\bibitem[{\citenamefont{Zar{\'a}nd et~al.}(2006)\citenamefont{Zar{\'a}nd,
  Chung, Simon, and Vojta}}]{zarand2006}
\bibinfo{author}{\bibfnamefont{G.}~\bibnamefont{Zar{\'a}nd}},
  \bibinfo{author}{\bibfnamefont{C.-H.} \bibnamefont{Chung}},
  \bibinfo{author}{\bibfnamefont{P.}~\bibnamefont{Simon}}, \bibnamefont{and}
  \bibinfo{author}{\bibfnamefont{M.}~\bibnamefont{Vojta}},
  \bibinfo{journal}{Phys. Rev. Lett.} \textbf{\bibinfo{volume}{97}},
  \bibinfo{pages}{166802} (\bibinfo{year}{2006}).

\bibitem[{\citenamefont{Maciejko et~al.}(2009)\citenamefont{Maciejko, Liu,
  Oreg, Qi, Wu, and Zhang}}]{maciejko2009}
\bibinfo{author}{\bibfnamefont{J.}~\bibnamefont{Maciejko}},
  \bibinfo{author}{\bibfnamefont{C.}~\bibnamefont{Liu}},
  \bibinfo{author}{\bibfnamefont{Y.}~\bibnamefont{Oreg}},
  \bibinfo{author}{\bibfnamefont{X.-L.} \bibnamefont{Qi}},
  \bibinfo{author}{\bibfnamefont{C.}~\bibnamefont{Wu}}, \bibnamefont{and}
  \bibinfo{author}{\bibfnamefont{S.-C.} \bibnamefont{Zhang}},
  \bibinfo{journal}{Phys. Rev. Lett.} \textbf{\bibinfo{volume}{102}},
  \bibinfo{pages}{256803} (\bibinfo{year}{2009}).

\bibitem[{\citenamefont{K\"onig}(2007)}]{konigphd}
\bibinfo{author}{\bibfnamefont{M.}~\bibnamefont{K\"onig}}, Ph.D. thesis,
  \bibinfo{school}{University of W\"urzburg} (\bibinfo{year}{2007}).

\bibitem[{\citenamefont{Law et~al.}(2010)\citenamefont{Law, Seng, Lee, and
  Ng}}]{law2010}
\bibinfo{author}{\bibfnamefont{K.~T.} \bibnamefont{Law}},
  \bibinfo{author}{\bibfnamefont{C.~Y.} \bibnamefont{Seng}},
  \bibinfo{author}{\bibfnamefont{P.~A.} \bibnamefont{Lee}}, \bibnamefont{and}
  \bibinfo{author}{\bibfnamefont{T.~K.} \bibnamefont{Ng}},
  \bibinfo{journal}{Phys. Rev. Lett.} \textbf{\bibinfo{volume}{81}},
  \bibinfo{pages}{041305(R)} (\bibinfo{year}{2010}).

\bibitem[{\citenamefont{Shindou et~al.}(2010)\citenamefont{Shindou, Furusaki,
  and Nagaosa}}]{shindou2010}
\bibinfo{author}{\bibfnamefont{R.}~\bibnamefont{Shindou}},
  \bibinfo{author}{\bibfnamefont{A.}~\bibnamefont{Furusaki}}, \bibnamefont{and}
  \bibinfo{author}{\bibfnamefont{N.}~\bibnamefont{Nagaosa}},
  \bibinfo{journal}{Phys. Rev. B} \textbf{\bibinfo{volume}{82}},
  \bibinfo{pages}{180505(R)} (\bibinfo{year}{2010}).

\bibitem[{\citenamefont{Qi et~al.}(2010)\citenamefont{Qi, Hughes, and
  Zhang}}]{qi2010tsc}
\bibinfo{author}{\bibfnamefont{X.-L.} \bibnamefont{Qi}},
  \bibinfo{author}{\bibfnamefont{T.~L.} \bibnamefont{Hughes}},
  \bibnamefont{and} \bibinfo{author}{\bibfnamefont{S.-C.} \bibnamefont{Zhang}},
  \bibinfo{journal}{Phys. Rev. B} \textbf{\bibinfo{volume}{82}},
  \bibinfo{pages}{184516} (\bibinfo{year}{2010}).

\bibitem[{\citenamefont{Liu et~al.}(2008)\citenamefont{Liu, Qi, Dai, Fang, and
  Zhang}}]{liu2008qah}
\bibinfo{author}{\bibfnamefont{C.-X.} \bibnamefont{Liu}},
  \bibinfo{author}{\bibfnamefont{X.-L.} \bibnamefont{Qi}},
  \bibinfo{author}{\bibfnamefont{X.}~\bibnamefont{Dai}},
  \bibinfo{author}{\bibfnamefont{Z.}~\bibnamefont{Fang}}, \bibnamefont{and}
  \bibinfo{author}{\bibfnamefont{S.-C.} \bibnamefont{Zhang}},
  \bibinfo{journal}{Phys. Rev. Lett.} \textbf{\bibinfo{volume}{101}},
  \bibinfo{pages}{146802} (\bibinfo{year}{2008}).

\bibitem[{\citenamefont{Yu et~al.}(2010)\citenamefont{Yu, Zhang, Zhang, Zhang,
  Dai, and Fang}}]{yu2010qah}
\bibinfo{author}{\bibfnamefont{R.}~\bibnamefont{Yu}},
  \bibinfo{author}{\bibfnamefont{W.}~\bibnamefont{Zhang}},
  \bibinfo{author}{\bibfnamefont{H.-J.} \bibnamefont{Zhang}},
  \bibinfo{author}{\bibfnamefont{S.-C.} \bibnamefont{Zhang}},
  \bibinfo{author}{\bibfnamefont{X.}~\bibnamefont{Dai}}, \bibnamefont{and}
  \bibinfo{author}{\bibfnamefont{Z.}~\bibnamefont{Fang}},
  \bibinfo{journal}{Science} \textbf{\bibinfo{volume}{329}},
  \bibinfo{pages}{61} (\bibinfo{year}{2010}).

\bibitem[{\citenamefont{Bolech and Demler}(2007)}]{bolech2007}
\bibinfo{author}{\bibfnamefont{C.~J.} \bibnamefont{Bolech}} \bibnamefont{and}
  \bibinfo{author}{\bibfnamefont{E.}~\bibnamefont{Demler}},
  \bibinfo{journal}{Phys. Rev. Lett.} \textbf{\bibinfo{volume}{98}},
  \bibinfo{pages}{237002} (\bibinfo{year}{2007}).

\bibitem[{\citenamefont{Coleman and Schofield}(1995)}]{coleman1995prl}
\bibinfo{author}{\bibfnamefont{P.}~\bibnamefont{Coleman}} \bibnamefont{and}
  \bibinfo{author}{\bibfnamefont{A.~J.} \bibnamefont{Schofield}},
  \bibinfo{journal}{Phys. Rev. Lett.} \textbf{\bibinfo{volume}{75}},
  \bibinfo{pages}{2184} (\bibinfo{year}{1995}).

\bibitem[{\citenamefont{Coleman et~al.}(1995)\citenamefont{Coleman, Ioffe, and
  Tsvelik}}]{coleman1995}
\bibinfo{author}{\bibfnamefont{P.}~\bibnamefont{Coleman}},
  \bibinfo{author}{\bibfnamefont{L.~B.} \bibnamefont{Ioffe}}, \bibnamefont{and}
  \bibinfo{author}{\bibfnamefont{A.~M.} \bibnamefont{Tsvelik}},
  \bibinfo{journal}{Phys. Rev. B} \textbf{\bibinfo{volume}{52}},
  \bibinfo{pages}{6611} (\bibinfo{year}{1995}).

\bibitem[{\citenamefont{Bradley et~al.}(1999)\citenamefont{Bradley, Bulla,
  Hewson, and Zhang}}]{bradley1999}
\bibinfo{author}{\bibfnamefont{S.~C.} \bibnamefont{Bradley}},
  \bibinfo{author}{\bibfnamefont{R.}~\bibnamefont{Bulla}},
  \bibinfo{author}{\bibfnamefont{A.~C.} \bibnamefont{Hewson}},
  \bibnamefont{and} \bibinfo{author}{\bibfnamefont{G.-M.} \bibnamefont{Zhang}},
  \bibinfo{journal}{Eur. Phys. J. B} \textbf{\bibinfo{volume}{11}},
  \bibinfo{pages}{535} (\bibinfo{year}{1999}).

\bibitem[{\citenamefont{Jones et~al.}(1988)\citenamefont{Jones, Varma, and
  Wilkins}}]{jones1988}
\bibinfo{author}{\bibfnamefont{B.~A.} \bibnamefont{Jones}},
  \bibinfo{author}{\bibfnamefont{C.~M.} \bibnamefont{Varma}}, \bibnamefont{and}
  \bibinfo{author}{\bibfnamefont{J.~W.} \bibnamefont{Wilkins}},
  \bibinfo{journal}{Phys. Rev. Lett.} \textbf{\bibinfo{volume}{61}},
  \bibinfo{pages}{125} (\bibinfo{year}{1988}).

\bibitem[{\citenamefont{Bolech and Iucci}(2006)}]{bolech2006}
\bibinfo{author}{\bibfnamefont{C.~J.} \bibnamefont{Bolech}} \bibnamefont{and}
  \bibinfo{author}{\bibfnamefont{A.}~\bibnamefont{Iucci}},
  \bibinfo{journal}{Phys. Rev. Lett.} \textbf{\bibinfo{volume}{96}},
  \bibinfo{pages}{056402} (\bibinfo{year}{2006}).

\bibitem[{\citenamefont{Martinek
  et~al.}(2003{\natexlab{a}})\citenamefont{Martinek, Utsumi, Imamura, Barnas,
  Maekawa, K\"onig, and Schon}}]{martinek2003a}
\bibinfo{author}{\bibfnamefont{J.}~\bibnamefont{Martinek}},
  \bibinfo{author}{\bibfnamefont{Y.}~\bibnamefont{Utsumi}},
  \bibinfo{author}{\bibfnamefont{H.}~\bibnamefont{Imamura}},
  \bibinfo{author}{\bibfnamefont{J.}~\bibnamefont{Barnas}},
  \bibinfo{author}{\bibfnamefont{S.}~\bibnamefont{Maekawa}},
  \bibinfo{author}{\bibfnamefont{J.}~\bibnamefont{K\"onig}}, \bibnamefont{and}
  \bibinfo{author}{\bibfnamefont{G.}~\bibnamefont{Schon}},
  \bibinfo{journal}{Phys. Rev. Lett.} \textbf{\bibinfo{volume}{91}},
  \bibinfo{pages}{127203} (\bibinfo{year}{2003}{\natexlab{a}}).

\bibitem[{\citenamefont{Martinek
  et~al.}(2003{\natexlab{b}})\citenamefont{Martinek, Sindel, Borda, Barna{\'s},
  K\"onig, Sch\"on, and von Delft}}]{martinek2003b}
\bibinfo{author}{\bibfnamefont{J.}~\bibnamefont{Martinek}},
  \bibinfo{author}{\bibfnamefont{M.}~\bibnamefont{Sindel}},
  \bibinfo{author}{\bibfnamefont{L.}~\bibnamefont{Borda}},
  \bibinfo{author}{\bibfnamefont{J.}~\bibnamefont{Barna{\'s}}},
  \bibinfo{author}{\bibfnamefont{J.}~\bibnamefont{K\"onig}},
  \bibinfo{author}{\bibfnamefont{G.}~\bibnamefont{Sch\"on}}, \bibnamefont{and}
  \bibinfo{author}{\bibfnamefont{J.}~\bibnamefont{von Delft}},
  \bibinfo{journal}{Phys. Rev. Lett.} \textbf{\bibinfo{volume}{91}},
  \bibinfo{pages}{247202} (\bibinfo{year}{2003}{\natexlab{b}}).

\bibitem[{\citenamefont{Choi et~al.}(2004)\citenamefont{Choi, Sanchez, and
  L\'opez}}]{choi2004}
\bibinfo{author}{\bibfnamefont{M.-S.} \bibnamefont{Choi}},
  \bibinfo{author}{\bibfnamefont{D.}~\bibnamefont{Sanchez}}, \bibnamefont{and}
  \bibinfo{author}{\bibfnamefont{R.}~\bibnamefont{L\'opez}},
  \bibinfo{journal}{Phys. Rev. Lett.} \textbf{\bibinfo{volume}{92}},
  \bibinfo{pages}{056601} (\bibinfo{year}{2004}).

\bibitem[{\citenamefont{Wilson}(1975)}]{wilson1975}
\bibinfo{author}{\bibfnamefont{K.~G.} \bibnamefont{Wilson}},
  \bibinfo{journal}{Rev. Mod. Phys.} \textbf{\bibinfo{volume}{47}},
  \bibinfo{pages}{773} (\bibinfo{year}{1975}).

\bibitem[{\citenamefont{Krishna-murthy
  et~al.}(1980{\natexlab{a}})\citenamefont{Krishna-murthy, Wilkins, and
  Wilson}}]{krishna1980a}
\bibinfo{author}{\bibfnamefont{H.~R.} \bibnamefont{Krishna-murthy}},
  \bibinfo{author}{\bibfnamefont{J.~W.} \bibnamefont{Wilkins}},
  \bibnamefont{and} \bibinfo{author}{\bibfnamefont{K.~G.}
  \bibnamefont{Wilson}}, \bibinfo{journal}{Phys. Rev. B}
  \textbf{\bibinfo{volume}{21}}, \bibinfo{pages}{1003}
  (\bibinfo{year}{1980}{\natexlab{a}}).

\bibitem[{\citenamefont{Krishna-murthy
  et~al.}(1980{\natexlab{b}})\citenamefont{Krishna-murthy, Wilkins, and
  Wilson}}]{krishna1980b}
\bibinfo{author}{\bibfnamefont{H.~R.} \bibnamefont{Krishna-murthy}},
  \bibinfo{author}{\bibfnamefont{J.~W.} \bibnamefont{Wilkins}},
  \bibnamefont{and} \bibinfo{author}{\bibfnamefont{K.~G.}
  \bibnamefont{Wilson}}, \bibinfo{journal}{Phys. Rev. B}
  \textbf{\bibinfo{volume}{21}}, \bibinfo{pages}{1044}
  (\bibinfo{year}{1980}{\natexlab{b}}).

\bibitem[{\citenamefont{Bulla et~al.}(2008)\citenamefont{Bulla, Costi, and
  Pruschke}}]{bulla2008}
\bibinfo{author}{\bibfnamefont{R.}~\bibnamefont{Bulla}},
  \bibinfo{author}{\bibfnamefont{T.}~\bibnamefont{Costi}}, \bibnamefont{and}
  \bibinfo{author}{\bibfnamefont{T.}~\bibnamefont{Pruschke}},
  \bibinfo{journal}{Rev. Mod. Phys.} \textbf{\bibinfo{volume}{80}},
  \bibinfo{pages}{395} (\bibinfo{year}{2008}).

\bibitem[{\citenamefont{Oliveira and Oliveira}(1994)}]{oliveira1994}
\bibinfo{author}{\bibfnamefont{W.~C.} \bibnamefont{Oliveira}} \bibnamefont{and}
  \bibinfo{author}{\bibfnamefont{L.~N.} \bibnamefont{Oliveira}},
  \bibinfo{journal}{Phys. Rev. B} \textbf{\bibinfo{volume}{49}},
  \bibinfo{pages}{11986} (\bibinfo{year}{1994}).

\bibitem[{\citenamefont{Campo and Oliveira}(2005)}]{campo2005}
\bibinfo{author}{\bibfnamefont{V.~L.} \bibnamefont{Campo}} \bibnamefont{and}
  \bibinfo{author}{\bibfnamefont{L.~N.} \bibnamefont{Oliveira}},
  \bibinfo{journal}{Phys. Rev. B} \textbf{\bibinfo{volume}{72}},
  \bibinfo{pages}{104432} (\bibinfo{year}{2005}).

\bibitem[{\citenamefont{\v{Z}itko and Pruschke}(2009)}]{resolution}
\bibinfo{author}{\bibfnamefont{R.}~\bibnamefont{\v{Z}itko}} \bibnamefont{and}
  \bibinfo{author}{\bibfnamefont{T.}~\bibnamefont{Pruschke}},
  \bibinfo{journal}{Phys. Rev. B} \textbf{\bibinfo{volume}{79}},
  \bibinfo{pages}{085106} (\bibinfo{year}{2009}).

\bibitem[{\citenamefont{Hofstetter}(2000)}]{hofstetter2000}
\bibinfo{author}{\bibfnamefont{W.}~\bibnamefont{Hofstetter}},
  \bibinfo{journal}{Phys. Rev. Lett.} \textbf{\bibinfo{volume}{85}},
  \bibinfo{pages}{1508} (\bibinfo{year}{2000}).

\bibitem[{\citenamefont{Peters et~al.}(2006)\citenamefont{Peters, Pruschke, and
  Anders}}]{peters2006}
\bibinfo{author}{\bibfnamefont{R.}~\bibnamefont{Peters}},
  \bibinfo{author}{\bibfnamefont{T.}~\bibnamefont{Pruschke}}, \bibnamefont{and}
  \bibinfo{author}{\bibfnamefont{F.~B.} \bibnamefont{Anders}},
  \bibinfo{journal}{Phys. Rev. B} \textbf{\bibinfo{volume}{74}},
  \bibinfo{pages}{245114} (\bibinfo{year}{2006}).

\bibitem[{\citenamefont{Weichselbaum and von Delft}(2007)}]{weichselbaum2007}
\bibinfo{author}{\bibfnamefont{A.}~\bibnamefont{Weichselbaum}}
  \bibnamefont{and} \bibinfo{author}{\bibfnamefont{J.}~\bibnamefont{von
  Delft}}, \bibinfo{journal}{Phys. Rev. Lett.} \textbf{\bibinfo{volume}{99}},
  \bibinfo{pages}{076402} (\bibinfo{year}{2007}).

\bibitem[{\citenamefont{Costi}(2001)}]{costi2001}
\bibinfo{author}{\bibfnamefont{T.~A.} \bibnamefont{Costi}},
  \bibinfo{journal}{Phys. Rev. B} \textbf{\bibinfo{volume}{64}},
  \bibinfo{pages}{241310(R)} (\bibinfo{year}{2001}).

\bibitem[{\citenamefont{Yoshida et~al.}(2009)\citenamefont{Yoshida, Seridonio,
  and Oliveira}}]{yoshida2009prb}
\bibinfo{author}{\bibfnamefont{M.}~\bibnamefont{Yoshida}},
  \bibinfo{author}{\bibfnamefont{A.~C.} \bibnamefont{Seridonio}},
  \bibnamefont{and} \bibinfo{author}{\bibfnamefont{L.~N.}
  \bibnamefont{Oliveira}}, \bibinfo{journal}{Phys. Rev. B}
  \textbf{\bibinfo{volume}{80}}, \bibinfo{pages}{235317}
  (\bibinfo{year}{2009}).

\bibitem[{\citenamefont{Meir and Wingreen}(1992)}]{meir1992}
\bibinfo{author}{\bibfnamefont{Y.}~\bibnamefont{Meir}} \bibnamefont{and}
  \bibinfo{author}{\bibfnamefont{N.~S.} \bibnamefont{Wingreen}},
  \bibinfo{journal}{Phys. Rev. Lett.} \textbf{\bibinfo{volume}{68}},
  \bibinfo{pages}{2512} (\bibinfo{year}{1992}).

\bibitem[{\citenamefont{Haldane}(1978)}]{haldane1978}
\bibinfo{author}{\bibfnamefont{F.~D.~M.} \bibnamefont{Haldane}},
  \bibinfo{journal}{Phys. Rev. Lett.} \textbf{\bibinfo{volume}{40}},
  \bibinfo{pages}{416} (\bibinfo{year}{1978}).

\bibitem[{\citenamefont{Hewson}(1993)}]{hewson}
\bibinfo{author}{\bibfnamefont{A.~C.} \bibnamefont{Hewson}},
  \emph{\bibinfo{title}{The Kondo Problem to Heavy-Fermions}}
  (\bibinfo{publisher}{Cambridge University Press, Cambridge},
  \bibinfo{year}{1993}).

\bibitem[{\citenamefont{Goldhaber-Gordon
  et~al.}(1998)\citenamefont{Goldhaber-Gordon, G\"ores, Kastner, Shtrikman,
  Mahalu, and Meirav}}]{goldhabergordon1998a}
\bibinfo{author}{\bibfnamefont{D.}~\bibnamefont{Goldhaber-Gordon}},
  \bibinfo{author}{\bibfnamefont{J.}~\bibnamefont{G\"ores}},
  \bibinfo{author}{\bibfnamefont{M.~A.} \bibnamefont{Kastner}},
  \bibinfo{author}{\bibfnamefont{H.}~\bibnamefont{Shtrikman}},
  \bibinfo{author}{\bibfnamefont{D.}~\bibnamefont{Mahalu}}, \bibnamefont{and}
  \bibinfo{author}{\bibfnamefont{U.}~\bibnamefont{Meirav}},
  \bibinfo{journal}{Phys. Rev. Lett.} \textbf{\bibinfo{volume}{81}},
  \bibinfo{pages}{5225} (\bibinfo{year}{1998}).

\bibitem[{\citenamefont{Parks et~al.}(2010)\citenamefont{Parks, Champagne,
  Costi, Shum, Pasupathy, Neuscamman, Flores-Torres, Cornaglia, Aligia,
  Balseiro et~al.}}]{parks2010}
\bibinfo{author}{\bibfnamefont{J.~J.} \bibnamefont{Parks}},
  \bibinfo{author}{\bibfnamefont{A.~R.} \bibnamefont{Champagne}},
  \bibinfo{author}{\bibfnamefont{T.~A.} \bibnamefont{Costi}},
  \bibinfo{author}{\bibfnamefont{W.~W.} \bibnamefont{Shum}},
  \bibinfo{author}{\bibfnamefont{A.~N.} \bibnamefont{Pasupathy}},
  \bibinfo{author}{\bibfnamefont{E.}~\bibnamefont{Neuscamman}},
  \bibinfo{author}{\bibfnamefont{S.}~\bibnamefont{Flores-Torres}},
  \bibinfo{author}{\bibfnamefont{P.~S.} \bibnamefont{Cornaglia}},
  \bibinfo{author}{\bibfnamefont{A.~A.} \bibnamefont{Aligia}},
  \bibinfo{author}{\bibfnamefont{C.~A.} \bibnamefont{Balseiro}},
  \bibnamefont{et~al.}, \bibinfo{journal}{Science}
  \textbf{\bibinfo{volume}{328}}, \bibinfo{pages}{1370} (\bibinfo{year}{2010}).

\bibitem[{\citenamefont{Pasupathy et~al.}(2004)\citenamefont{Pasupathy,
  Bialczak, Martinek, Grose, Donev, McEuer, and Ralph}}]{pasupathy2004}
\bibinfo{author}{\bibfnamefont{A.~N.} \bibnamefont{Pasupathy}},
  \bibinfo{author}{\bibfnamefont{R.~C.} \bibnamefont{Bialczak}},
  \bibinfo{author}{\bibfnamefont{J.}~\bibnamefont{Martinek}},
  \bibinfo{author}{\bibfnamefont{J.~E.} \bibnamefont{Grose}},
  \bibinfo{author}{\bibfnamefont{L.~A.~K.} \bibnamefont{Donev}},
  \bibinfo{author}{\bibfnamefont{P.~L.} \bibnamefont{McEuer}},
  \bibnamefont{and} \bibinfo{author}{\bibfnamefont{D.~C.} \bibnamefont{Ralph}},
  \bibinfo{journal}{Science} \textbf{\bibinfo{volume}{306}},
  \bibinfo{pages}{86} (\bibinfo{year}{2004}).

\bibitem[{\citenamefont{Costi}(2000)}]{costi2000}
\bibinfo{author}{\bibfnamefont{T.~A.} \bibnamefont{Costi}},
  \bibinfo{journal}{Phys. Rev. Lett.} \textbf{\bibinfo{volume}{85}},
  \bibinfo{pages}{1504} (\bibinfo{year}{2000}).

\end{thebibliography}

\end{document}